\documentclass[twoside]{article}

\usepackage{lipsum} 

\usepackage{graphics,graphicx} 

\usepackage[sc]{mathpazo} 
\usepackage[T1]{fontenc} 
\usepackage{microtype} 

\usepackage[hmarginratio=1:1,top=32mm,columnsep=20pt]{geometry} 
\usepackage[hang, small,labelfont=bf,up,textfont=it,up]{caption} 
\usepackage{booktabs} 
\usepackage{float} 

\usepackage{lettrine} 
\usepackage{paralist} 

\usepackage{amsmath}

\usepackage{color}

\usepackage{subcaption}

\usepackage{abstract} 

\usepackage{titlesec} 

\usepackage{fancyhdr} 
\usepackage{placeins}
\providecommand{\keywords}[1]{\textbf{Keywords:} #1}

\title{\vspace{-15mm}\fontsize{16pt}{10pt}\selectfont\textbf{Spatio-temporal analysis of  regional unemployment rates: A comparison of model based approaches}} 

\author{
\large
Soraia Pereira$^1$, Feridun Turkman$^2$, Luis Correia$^3$
\\[2mm] 
\normalsize $^1$Universidade de Lisboa, $^2$Universidade de Lisboa, $^3$Instituto Nacional de Estatistica, Portugal \\ 
\vspace{-5mm}
}
\date{}

\begin{document}

\maketitle 

\begin{abstract}
\normalfont {\noindent This study aims to analyze the methodologies that can be used to estimate the total number of unemployed, as well as  the unemployment rates for 28 regions of Portugal, designated as NUTS III regions,  using model based approaches as compared to the direct estimation methods currently employed by INE (National Statistical Institute of Portugal). Model based methods, often known as small area estimation methods (Rao, 2003),  "borrow strength" from neighbouring  regions and in doing so, aim to compensate for the small sample sizes  often observed  in these areas. Consequently,   it is generally accepted that model based methods tend to produce estimates which have lesser variation.  Other   benefit in employing model based methods is the possibility of including   auxiliary information in the form of variables of interest and latent random structures. This study focuses on the application of Bayesian hierarchical models to the Portuguese Labor Force Survey data from the 1st quarter of 2011 to the 4th quarter of 2013. Three different data modeling strategies are considered and compared: Modeling of the total  unemployed through Poisson, Binomial and Negative Binomial models; modeling of rates using a Beta model; and modeling of the three states of the labor market (employed, unemployed and inactive) by a Multinomial model. The implementation of these models is based on the \textit{Integrated Nested Laplace Approximation} (INLA) approach, except for the Multinomial model which is implemented based on the method of Monte Carlo Markov Chain (MCMC). Finally,  a comparison of  the performance of these  models, as well as  the  comparison  of the results  with those obtained by direct estimation methods  at NUTS III level are given.
\\

\noindent \keywords{Unemployment estimation; Model based methods; Bayesian hierarchical models}}

\end{abstract}



\section{Introduction}

The calculation of official estimates of the labor market that are published quarterly by the INE is based on a direct method from the sample of the Portuguese Labor Force Survey. These estimates are available at national level and NUTS II regions of Portugal. NUTS is the classification of territorial units for statistics (see Appendix for a better understanding). Currently, as established by Eurostat, knowledge of the labor market requires reliable estimates for the total of unemployed people and the unemployment rate at more disaggregated levels, particularly at NUTS III level. However, due to the small size of these areas, there is insufficient information on some of the variables of interest to obtain estimates with acceptable accuracy using the direct method.

In this sense, and because increasing the sample size  imposes excessive costs, we intend to study alternative methods with the aim of getting more accurate estimates for these regions. In fact, the accuracy of the estimates obtained in this context is  deemed very important since it  directly  affects the local policy actions.

This issue is part of the \textit{small area estimation}. Rao (2003) provides a good theoretical introduction to this problem and discusses some estimation techniques based on mixed generalized models. Pfeffermann (2002), Jiang \& Lahiri (2006a, 2006b) make a good review of developments to date.

This has been an area in full development and application, especially after the incorporation of spatial and temporal random effects, which brought a major improvement in the estimates produced. Choundry \& Rao (1989), Rao \& Yu (1994), Singh et al (2005) and Lopez-Vizcaino et al (2015) are responsible for some of these developments. Chambers {\it et al} (2016)  give alternative semiparametric  methods based on M-quantile regression.

Datta \& Ghosh (1991) use a Bayesian approach for the estimation in small areas. One advantage of using this approach is the flexibility in modeling different types of variables of interest and different structures in the random effects using the same computational methods.

Recently, there has been considerable developments on space-time  Bayesian  hierarchical models employed in  small area estimation  within the context of disease (Best \textit{et al}, 2005). In this paper, we explore the application of these models and adopt them for  the estimation of unemployment in the NUTS III regions, using data from the Portuguese Labor Force Survey from the 1st quarter of 2011 to the 4th quarter of 2013.

We consider three different modeling strategies: the modeling of the total number of unemployed people through the Poisson, Binomial, and Negative Binomial models; modeling the unemployment rate using a Beta model; and the simultaneous modeling of the total of the three categories of the labor market (employment, unemployment and inactivity) using a Multinomial model.


\section{The data}

The region under study (Portugal Continental) is partitioned into 28 NUTS III regions, indexed by $ j = 1, ..., 28$. We did not include the autonomous regions because they coincide with the NUTS II regions for which estimates are already available with acceptable accuracy.

We use the Portuguese Labor Force Survey data from the 1st quarter of 2011 to the 4th quarter of 2013 in order to produce accurate estimates for the labor market indicators in the last quarter. Each quarter is denoted by $ t = 1, ..., 12 $. We did not use more recent data because there was a change in the sampling design during 2014 and that could affect the temporal analysis.

We are interested in the total unemployed population, and the unemployment rate of the population by NUTS III regions, which is denoted by $ Y_{jt} $ and $ R_{jt} $. We denote the respective sample values by $ y_{jt} $ and $ r_{jt} $. The unemployment rate is given by the ratio of active people who are unemployed, as defined by the European regulation of the Labor Force Survey.

The models developed to make estimation in small areas gain special importance with the inclusion of variables of interest, which we call  covariates. In this study, the covariates are divided into 5 groups: population structure, economy, labor market, companies and type of economic activity. Some of these covariates  are regional and are static in time  whereas others are available per quarter and thus are also of dynamic nature.  We will make the distinction and classify these  sets of covariates into  regional, temporal and spatio-temporal covariates.  These selected covariates are as follows:

\begin{enumerate}
\item Population structure: a) Proportion of individuals in the sample of the Labor Force Survey that are female and aged between 24 and 34 years (SA6, regional and quarterly); b) Proportion of individuals in the sample of the Labor Force Survey that are female and over 49 years (SA8, regional and quarterly).
\item Economy: a) Gross domestic product per capita (GDP, quarterly).
 \item Labor market: a) Proportion of unemployed people registered in the employment centers (IEFP, regional and quarterly).
\item Companies: a) Number of enterprises per 100 inhabitants (regional).
\item Type of economic activity: a) Proportion of population employed in the primary sector of activity (regional); b) Proportion of population employed in the secondary sector of activity (regional).
\end{enumerate}

\begin{figure}
\centering
\includegraphics[width=7cm]{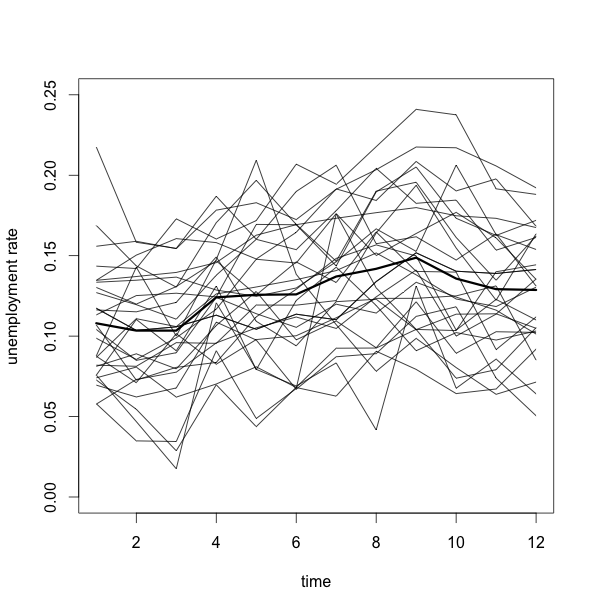}
\caption{Unemployment rate observed in the sample from the Portuguese Labor Force Survey from the 1st quarter of 2011 to the 4th quarter of 2013 in each of the 28 NUTS III}
\label{tdamostra}
\end{figure}

Figure \ref{tdamostra} shows the evolution of the unemployment rate observed in the sample from the Portuguese Labor Force Survey from the 1st quarter of 2011 to the 4th quarter of 2013 in each of the 28 NUTS III . The bold represents the average unemployment rate. We can see that for all regions there was a slight increase in the unemployment rate during this period.

\begin{figure}
\centering
\includegraphics[width=10cm]{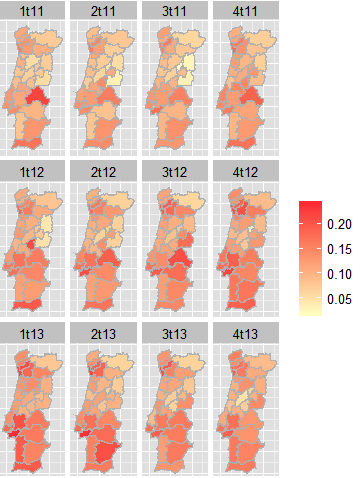}
\caption{Spatial and temporal distribution of the unemployment rate observed in the sample of Portuguese Labor Force Survey.}
\label{mapa_tdamostra}
\end{figure}

The map in Figure \ref{mapa_tdamostra}  shows the spatial and temporal distribution of the unemployment rate observed in the sample of Portuguese Labor Force Survey during the period under study. As we can see, this map suggests  the existence of spatial and temporal dependence structures in the observed data.


\FloatBarrier

\section{Bayesian models for counts and proportions}

In this problem we are interested in estimating the effect of selected variables on the  number of unemployed individuals and the unemployment rate, taking into account the temporal and spatial correlations.

One of the most general and useful  ways  of specifiying this problem is to employ   hierarchical generalized linear model set up, in which the data  are linked to covariates and spatial-temporal random effects through an appropriately chosen likelihood and a link  function  which is linear on the covariates and the random effects. 

We denote the vector of  designated regional covariates by $ \boldsymbol{x_j} = (x_{1j}, x_{2j}, x_{3j}) $, the temporal  covariates by  $ x_t $ and the vector of spatio-temporal covariates by $ \boldsymbol{x_{jt}} = (x_{1jt}, x_{2jt}, x_{3jt}) $.

While modeling unemployment numbers, we generically assume that

\begin{equation*}
y_{jt}|\mu_{jt} \sim \pi(y_{jt}|\mu_{jt}), \quad  j=1,...,28, \quad t=1,...,12,
\end{equation*}

\noindent where $ \pi $ is a generic probability mass  function. We look at this model considering specific probability mass functions, such as Poisson and Binomial, among others. The state parameters $ \mu_{jt} $ depend on covariates and on structured and unstructured random factors through appropriate link functions.

The unemployment rate is  also hierarchically modeled in a similar way. We assume that

\begin{equation*}
r_{jt}|\theta_{jt} \sim g(r_{jt}|\theta_{jt}) , \quad  j=1,...,28, \quad t=1,...,12,
\end{equation*}

\noindent where $ g $ is a properly chosen probability density function and $ \theta_{jt} $ are the state parameters.

In the following sections we look at different variations of these hierarchical structures with different link functions.

Let us consider $ h $,  the chosen link function which depends on the assumed model for the data. We assume $ \eta_{jt} = h (\mu_{jt}) $ for the modeling of the total and $ \eta_{jt} = h (\theta_{jt}) $ for the modeling of the rates. For each model, we consider the following linear predictor

\begin{equation}
\eta_{jt}=offset_{jt}+\alpha_0 + \boldsymbol{x'_j} \boldsymbol{\alpha} + x'_t \beta + \boldsymbol{x'_{jt}} \boldsymbol{\gamma} + w_{jt} + \epsilon_{jt}, \quad j=1,...,28 \quad t=1,...,12,
\end{equation}
\noindent where $offset_{jt}$ are constants that can be included in the linear predictor during adjustment. The vectors $ \boldsymbol{\alpha} = (\alpha_1, \alpha_2, \alpha_3) $, $ \beta $ and $ \boldsymbol{\gamma} = (\gamma_1, \gamma_2, \gamma_3) $ correspond respectively to vectors of the covariates coefficients $ \boldsymbol{x_j}$, $ x_t $ and $ \boldsymbol{x_{jt}} $. Components $ \epsilon_{jt} $ represent unstructured random effects, which assume

\begin{equation*}
\epsilon_{jt} \sim N(0, \sigma^2_{\epsilon}),
\end{equation*}
\noindent and the components $w_{jt}$ represent the structured random effects that can be written as $ w_{jt} = w_{1j} + w_{2t} $ where $ \boldsymbol{w_1} $ is modeled as a \textit{intrinsic conditional autoregressive} (ICAR) process proposed by Besag et al (1991) and $ \boldsymbol{w_2} $ is modeled as a first order \textit{random walk} (AR (1)). Blangiardo et al (2013) succinctly describe both the ICAR and AR (1) processes.

\begin{flalign*}
&\boldsymbol{w_1}|\tau_{w_1} \sim ICAR(\tau_{w_1}), \\
&\boldsymbol{w_2}|\tau_{w_2} \sim AR(1).
\end{flalign*}

\noindent We assume the following prior distributions for the regression parameters

\begin{flalign*}
&\alpha_0 \sim N(0,10^6), \\
&\alpha_i \sim N(0, 10^6) \ \quad i=1,2,3,\\
&\beta \sim N(0, 10^6), \\
&\gamma_i \sim N(0, 10^6)  \ \quad i=1,2,3. 
\end{flalign*}

\noindent For the hyperparameters we assume

\begin{flalign*}
&log \tau_{\epsilon} \sim log Gamma (1,0.0005), \\
&log \tau_{w_1} \sim log Gamma (1,0.0005), \\
&log \tau_{w_2} \sim log Gamma (1,0.0005).
\end{flalign*}

\noindent We assume the following models for the distribution of the observed data: Poisson, Binomial, and Negative Binomial for the total of unemployed, Beta for the unemployment rate and Multinomial for the total of the three states of the labor market (employment, unemployment and inactivity).

\subsection{Poisson model}

This is perhaps the most frequently used model for counting data in small areas, especially in epidemiology. If we consider that $ \mu_{jt} $ is the mean of the total number of unemployed people, we can assume that

\begin{equation*}
y_{jt}|\mu_{jt} \sim Poisson(\mu_{jt}), \ \quad j=1,...,28, \ \quad t=1,...,12.
\end{equation*}

\noindent Therefore

\begin{equation*}
p(y_{jt}|\mu_{jt})= \mu_{jt}^{y_{jt}} exp(-\mu_{jt})/ y_{jt}! , \quad y_{jt}=0,1,2...
\end{equation*}

\noindent In this case, the link function is the logarithmic function (log = $ h $). The NUTS III regions have different sample dimensions, so the variation of the total unemployment is affected. To remove this effect, we need to add an $offset$ term, which is given by the number of individuals in the sample in each NUTS III region.

\subsection{Negative Binomial model}

The Negative Binomial model may be used as an alternative to the Poisson model, especially when the sample variance is much higher than the sample mean. When this happens, we say that there is over-dispersion in the data.
In this case, we can assume that

\begin{equation*}
y_{jt}|\mu_{jt},\phi \sim \textit{Negative Binomial} (\mu_{jt},\phi), \ \quad j=1,...,28 \ \quad t=1,...,12.
\end{equation*}

\noindent The probability mass function is given by

\begin{equation*}
p(y_{jt}|\mu_{jt},\phi)= \frac{\Gamma(y_{jt}+\phi)}{\Gamma(\phi).y_{jt}!}. \frac{\mu_{jt}^{y_{jt}}. \phi^{\phi}}{(\mu_{jt}+\phi)^{y_{jt}+\phi}},  \quad y_{jt}=0,1,2...
\end{equation*}
\noindent where $\Gamma(.)$ is the gamma function.

The most convenient way to connect $ \mu_{jt} $ to the linear predictor is through the $ log\frac{\mu_{jt}}{\mu_{jt} +\phi}$. Also in this case, the term $offset$ described in the Poisson model is considered.

\subsection{Binomial model}

When measuring the total unemployed, we may also consider that there is a finite population in the area j. In this case, we assume that this population is the number of active individuals in the area j, which is denoted by $ m_{jt} $, assuming that it is fixed and known. We can then consider a Binomial model for the total number of unemployed given the observed active population. So, given the population unemployment rate $ R_{jt} $,

\begin{equation*}
y_{jt}|m_{jt},R_{jt} \sim Binomial(m_{jt},R_{jt}), \ \quad j=1,...,28 \ \quad t=1,...,12,
\end{equation*}

\noindent which means that

\begin{equation*}
p(y_{jt}|m_{jt},R_{jt})= {m_{jt} \choose y_{jt} } R_{jt}^{y_{jt}} (1-R_{jt})^{m_{jt}-y_{jt}},  \quad y_{jt}=0,1,...,m_{jt}, , \ \quad t=1,...,12.
\end{equation*}

\noindent In this case, the most usual link function is the logit function given by $ log (R_{jt} / (1-R_{jt})) $.

We expect that the fit of this model will be close to the fit of the Poisson model in the regions with a big number of active people and a small unemployment rate.


\subsection{Beta model}

The Beta distribution is one of the most commonly used  model for  rates and proportions. We can assume that the unemployment rate $ r_{jt} $ follows a Beta distribution and using the parameterization proposed by Ferrari and Cribari-Neto (2004), we denote by

\begin{equation*}
r_{jt}|\mu_{jt},\phi \sim Beta(\mu_{jt},\phi), \ \quad j=1,...,28 \ \quad t=1,...,12.
\end{equation*}

\noindent The probability mass function is given by

\begin{equation*}
p(r_{jt}|\mu_{jt},\phi)= \frac{\Gamma(\phi)}{\Gamma(\mu_{jt} \phi) \Gamma((1-\mu_{jt})\phi)} r_{jt}^{\mu_{jt} \phi -1} (1-r_{jt})^{(1-\mu_{jt})\phi-1}, \ \quad 0<r_{jt}<1,
\end{equation*}
\noindent where $0<\mu_{jt}<1$ and $\phi>0$.

In this case, there are several possible choices for the link function, but the most common is the logit function $ h (\mu_{jt}) = log (r_{jt} / (1-r_{jt})) $.

\subsection{Multinomial model}

The Multinomial logistic regression model is an extension of the Binomial logistic regression model and is used when the variable of interest is multi-category. In this case, it may interest us to model the three categories of the labor market (employment, unemployment and inactivity), giving us the unemployment rate which can be expressed by the ratio between the unemployed and the active people (the sum of the unemployed and employed).

One advantage of the Multinomial model in this problem is the consistency obtained between the three categories of the labor market. The estimated total employment, unemployment and inactivity coincides with the total population. In addition, the same model provides estimates for the rate of employment, unemployment and inactivity.

Assuming that $\boldsymbol{y_{jt}}=(y_{jt1},y_{jt2},y_{jt3})$ is the vector of the total in the three categories of the labor market, the Multinomial model can be written as

\begin{equation*}
\boldsymbol{y_{jt}}|n_{jt},\boldsymbol{P_{jt}} \sim Multinomial(n_{jt},\boldsymbol{P_{jt}}), \quad j=1,...,28 \quad t=1,...,12,
\end{equation*}
\noindent where $n_{jt} $ is the number of individuals in the area j and quarter t, and $\boldsymbol{P_{jt}}=(P_{jt1},P_{jt2},P_{jt3})$ is the vector of proportions of employed, unemployed and inactive, where $P_{jt3}=1-(P_{jt1}+P_{jt2})$.

\noindent The probability mass function is given by

\begin{equation*}
p(y_{jt1},y_{jt2},y_{jt3}|n_{jt},P_{jt})= \frac{n_{jt}!}{y_{jt1}!y_{jt2}!y_{jt3}!} P^{y_{jt1}}_{jt1} P^{y_{jt2}}_{jt2} P^{y_{jt3}}_{jt3} ,
\end{equation*}
where
\begin{equation*}
 y_{jtq} \in \mathbb{N} : \sum^{3}_q y_{jtq}=n_{jt}, \ q=1,2,3.
\end{equation*}

\noindent The most common link function is the log of $ P_{jtq} $, defined as $\eta_{jtq}=log(P_{jtq}/P_{jt3})$, $q=1,2$.


\FloatBarrier

\section{Application to the Portuguese Labor Force Survey data}

\subsection{Results}

This section provides the results of applying five models for the estimation of the total unemployed and unemployment rate to the NUTS III regions of Portugal.

The Poisson, Binomial, Negative Binomial and Beta models were implemented using the R package \textit{R-INLA}, while the Multinomial model was implemented based on MCMC methods using the R package \textit{R2OpenBUGS}. 

When the Multinomial regression model was combined with the predictor given in (1),  some convergence problems arose, due to its complexity. For this reason, the effects $ w_{jt} $ and $ \epsilon_{jt} $ were replaced by the unstructured area and time effects $ u_j $ and $ v_t $, where it was assumed

\begin{flalign*}
& u_j \sim N(0,\sigma^2_u),\\
& v_t \sim N(0,\sigma^2_v),
\end{flalign*}

\noindent with the following prior information

\begin{flalign*}
&log \tau_{u} \sim log Gamma (1,0.0005), \\
&log \tau_{v} \sim log Gamma (1,0.0005).
\end{flalign*}

\noindent Due to the differences in the model structure and the  computational methods used for  the Multinomial model,  the comparative analysis of results for this model  should be done with some extra care.

The posterior mean of the parameters and hyperparameters of each model as well as the standard deviation and the quantile 2.5 \% and 97.5 \% are presented in Tables \ref{coef_bin_todos}, \ref{coef_bin_todos_nb}, \ref{coef_bin_todos_bin}, \ref{coef_bin_todos_beta} and \ref{coef_bin_mult}. We can see that the covariates \textit{GDP} and \textit{secondary sector} are not significant for any of the models applied. However, the value obtained for \textit{Deviance Information Criterion} (DIC) increases considerably without the inclusion of these variables, so we decided to include them.
 
 We  observe that the IEFP is significant in all of the models applied, as expected.  The number of enterprises per 100 000 inhabitants has a negative effect on the increase of unemployment.
The population structure has also a significant effect. The proportion of individuals that are female and aged between 24 and 34 years has a positive effect on the increase of unemployment. On the other hand, the proportion of individuals that are female and over 49 years has a negative effect. These tendencies are probably due to young unemployment in the first case and to the fact that the age group +49 includes most of the inactive people, in the second case.

\begin{table}
\captionsetup{width=0.8\linewidth}
\caption{Posterior mean, standard deviation and 95 \% credibility interval for the parameters and hyperparameters of Poisson model. \label{coef_bin_todos}}
\centering
{\begin{tabular}{lrrrr}
\hline
&   \multicolumn{4}{c}{\textbf{Poisson}}  \\
\cline{2-5}
&  \textbf{Mean} & \textbf{SD} & \textbf{2.5Q} & \textbf{97.5Q}  \\
\hline
(Intercept)	&	-2,83	&	0,01	&	-2,85	&	-2,81	\\	
Companies	&	-0,01	&	0,02	&	-0,05	&	0,02	\\	
Primary	sector	&	-0,02	&	0,72	&	-1,45	&	1,40	\\
Secondary	sector	&	0,02	&	0,21	&	-0,39	&	0,43	\\
GDP	&	0,00	&	0,00	&	0,00	&	0,00	\\	
IEFP	&	10,05	&	0,96	&	8,17	&	11,93	\\	
SA6	&	4,30	&	1,34	&	1,65	&	6,93	\\	
SA8	&	-1,55	&	0,57	&	-2,65	&	-0,42	\\	
\hline										
$\tau$	&	&	&	&		\\	
$\tau_{w_2}$	&	25047,76	&	20819,39	&	3297,22	&	79744,21	\\	
$\tau_{w_1}$	&	25,77	&	9,52	&	11,91	&	48,78	\\	
$\tau_{\epsilon}$	&	22082,79	&	19692,19	&	2213,88	&	73957,91	\\	
\hline
\end{tabular}}
\end{table}

\begin{table}
\captionsetup{width=0.8\linewidth}
\caption{Posterior mean, standard deviation and 95 \% credibility interval for the parameters and hyperparameters of Negative Binomial model. \label{coef_bin_todos_nb}}
\centering
{\begin{tabular}{lrrrr}
\hline
&   \multicolumn{4}{c}{\textbf{Negative Binomial}}  \\
\cline{2-5}
&  \textbf{Mean} & \textbf{SD} & \textbf{2.5Q} & \textbf{97.5Q}  \\
\hline
(Intercept)	&	-2,83	&	0,01	&	-2,86	&	-2,81	\\	
Companies	&	-0,01	&	0,02	&	-0,05	&	0,02	\\	
Primary	sector	&	0,13	&	0,73	&	-1,33	&	1,57	\\
Secondary	sector	&	-0,04	&	0,23	&	-0,48	&	0,41	\\
GDP	&	0,00	&	0,00	&	0,00	&	0,00	\\	
IEFP	&	10,20	&	1,48	&	7,28	&	13,09	\\	
SA6	&	3,97	&	2,01	&	0,02	&	7,91	\\	
SA8	&	-2,11	&	0,73	&	-3,54	&	-0,67	\\	
\hline										
$\tau$	&	48,57	&	5,44	&	38,69	&	60,05	\\			
$\tau_{w_2}$	&	22946,14	&	20085,32	&	2453,14	&	75579,11	\\	
$\tau_{w_1}$	&	32,77	&	14,26	&	13,08	&	68,00	\\	
$\tau_{\epsilon}$	&	22641,11	&	19924,30	&	2405,53	&	74957,89	\\	
\hline
\end{tabular}}
\end{table}

\begin{table}
\captionsetup{width=0.8\linewidth}
\caption{Posterior mean, standard deviation and 95 \% credibility interval for the parameters and hyperparameters of Binomial model. \label{coef_bin_todos_bin}}
\centering
{\begin{tabular}{lrrrr}
\hline
&   \multicolumn{4}{c}{\textbf{Binomial}}  \\
\cline{2-5}
&  \textbf{Mean} & \textbf{SD} & \textbf{2.5Q} & \textbf{97.5Q}  \\
\hline
(Intercept)	&	-1,97	&	0,01	&	-2,00	&	-1,95	\\	
Companies	&	-0,04	&	0,02	&	-0,07	&	0,00	\\	
Primary	sector	&	0,54	&	1,01	&	-1,47	&	2,52	\\
Secondary	sector	&	-0,11	&	0,28	&	-0,67	&	0,45	\\
GDP	&	0,00	&	0,00	&	0,00	&	0,00	\\	
IEFP	&	12,63	&	1,11	&	10,47	&	14,81	\\	
SA6	&	4,38	&	1,47	&	1,50	&	7,26	\\	
SA8	&	-1,11	&	0,64	&	-2,37	&	0,16	\\	
\hline										
$\tau$	&	&	&	&					\\	
$\tau_{w_2}$	&	20736,79	&	19243,68	&	2070,46	&	71553,26	\\	
$\tau_{w_1}$	&	11,77	&	3,90	&	5,70	&	20,85	\\	
$\tau_{\epsilon}$	&	19143,06	&	18555,71	&	1460,60	&	68268,97	\\	
\hline
\end{tabular}}
\end{table}

\begin{table}
\captionsetup{width=0.8\linewidth}
\caption{Posterior mean, standard deviation and 95 \% credibility interval for the parameters and hyperparameters of Beta model. \label{coef_bin_todos_beta}}
\centering
{\begin{tabular}{lrrrr}
\hline
&   \multicolumn{4}{c}{\textbf{Beta}}  \\
\cline{2-5}
&  \textbf{Mean} & \textbf{SD} & \textbf{2.5Q} & \textbf{97.5Q}  \\
\hline
(Intercept)	&	-1,98	&	0,01	&	-2,00	&	-1,95	\\	
Companies	&	-0,03	&	0,03	&	-0,08	&	0,02	\\	
Primary	sector	&	0,69	&	1,09	&	-1,50	&	2,83	\\
Secondary	sector	&	-0,20	&	0,31	&	-0,82	&	0,42	\\
GDP	&	0,00	&	0,00	&	0,00	&	0,00	\\	
IEFP	&	12,22	&	1,82	&	8,64	&	15,78	\\	
SA6	&	0,85	&	1,84	&	-2,77	&	4,47	\\	
SA8	&	-2,37	&	0,68	&	-3,70	&	-1,03	\\	
\hline										
$\tau$	&	206,61	&	16,80	&	174,43	&	240,37	\\				
$\tau_{w_2}$	&	20012,58	&	19075,29	&	1750,73	&	70491,89	\\	
$\tau_{w_1}$	&	11,33	&	4,61	&	5,40	&	23,02	\\	
$\tau_{\epsilon}$	&	20497,48	&	19404,83	&	1715,97	&	71768,00	\\	
\hline										
\end{tabular}}
\end{table}

\begin{table}
\captionsetup{width=0.8\linewidth}
\caption{Posterior mean, standard deviation and 95 \% credibility interval for the parameters and hyperparameters of Multinomial model. \label{coef_bin_mult}}
\centering
{\begin{tabular}{lrrrr}
\hline
&   \multicolumn{4}{c}{\textbf{Multinomial}} \\
\cline{2-5}
&  \textbf{Mean} & \textbf{SD} & \textbf{2.5Q} & \textbf{97.5Q}  \\
\hline
(Intercpet)&	-1,74	&	0,26	&	-2,16	&	-1,20	\\		
Companies&	0,01	&	0,02	&	-0,04	&	0,05	\\		
Primary	sector	&	3,99	&	5,38	&	-1,04	&	14,94	\\
Secondary	sector	&	-0,77	&	0,77	&	-2,27	&	0,19	\\
GDP&	0,00	&	0,00	&	0,00	&	0,00	\\		
IEFP&	8,93	&	1,59	&	6,02	&	12,00	\\		
SA6&	4,77	&	1,44	&	1,97	&	7,58	\\		
SA8&	-2,38	&	0,64	&	-3,74	&	-1,22	\\		
\hline										
$\tau_v$	&	2206,25	&	2979,60	&	3,32	&	9519,00	\\	
$\tau_u$	&	33,57	&	24,87	&	1,77	&	78,11	\\	
\hline
\end{tabular}}
\end{table}


 All the considered models give very good fit to the data and their temporal predictions are also satisfactory. Here we report  on  several  model fitting  aspects  of  the Binomial model. Similar results for the other models are given in the Supplementary Material.  

Figure  \ref{ic_modelos_taxas} a) gives the observed and adjusted values from the Binomial model together with their 95\% credible intervals, whereas figure \ref{ic_modelos_taxas} b) gives the predictions to the 4th quarter of 2013 together with their 95\% credible intervals. We see that the adjusted values are very close to the observed ones. The domains are sorted at first by quarter and then by region. This is the reason for the identical behavior in each 28 domains (corresponding to the NUTS III regions).  The graphs show a slight increase on the unemployment rate until the 1st quarter of 2013 and then a decrease until the 4th quarter of 2013.

\begin{figure}[!htbp]
\begin{subfigure}{\linewidth}
  \centering
  \includegraphics[width=1\linewidth]{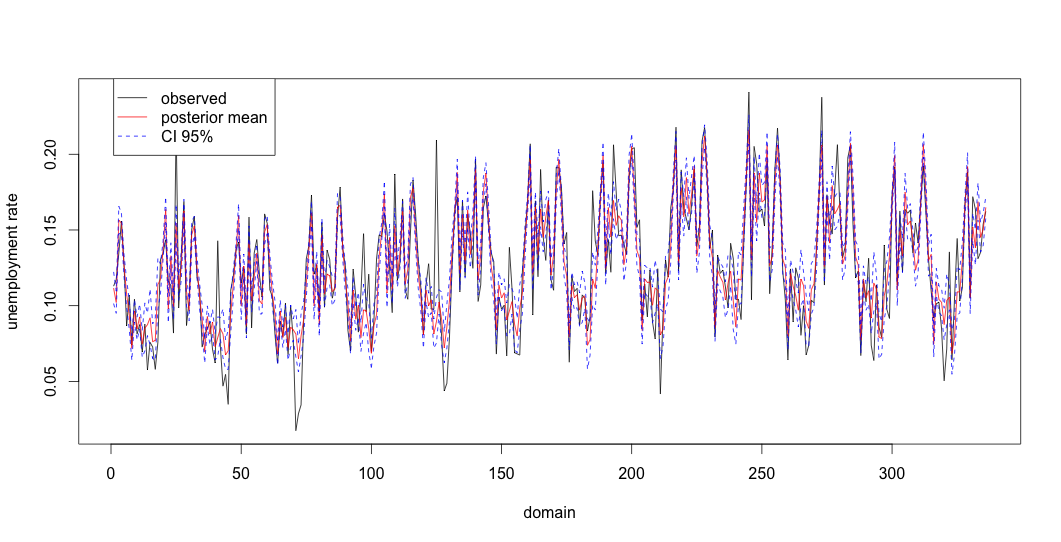}
  \caption{}
  \label{fig:sfig1}
\end{subfigure}%
\\
\begin{subfigure}{\linewidth}
  \centering
  \includegraphics[width=1\linewidth]{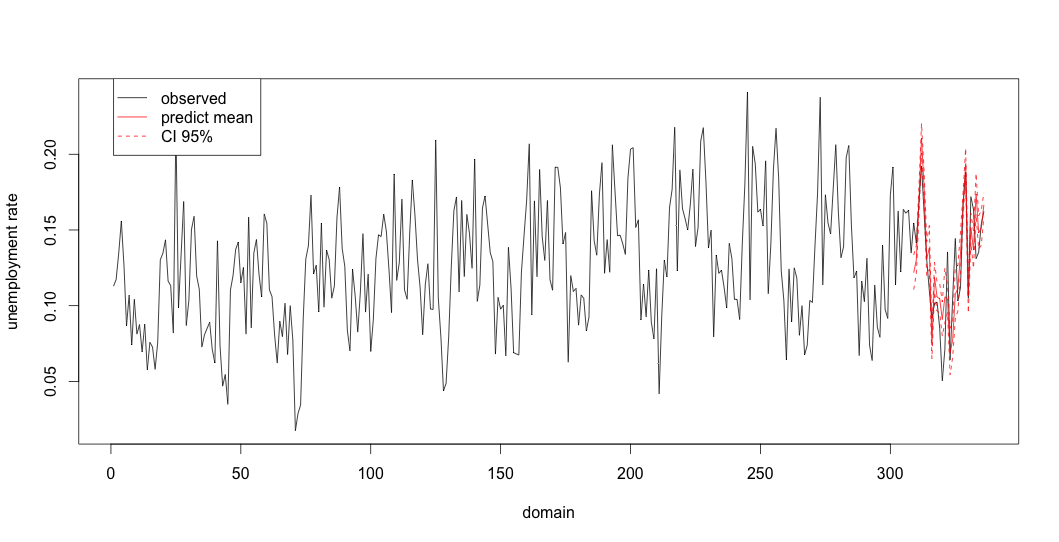}
 \caption{}
  \label{fig:sfig2}
\end{subfigure}
\caption{a) Observed and adjusted values (mean and 95 \% CI) of unemployment rate for the 336 domains (336 = 28 NUTS III $ \times $ 12 quarters); b) Observed and predicted values (the posterior mean and 95 \% CI) of the unemployment rate. The prediction is made for the 4th quarter of 2013 which is highlighted red.}
\label{ic_modelos_taxas}
\end{figure}



\begin{figure}
\centering
\includegraphics[width=14cm]{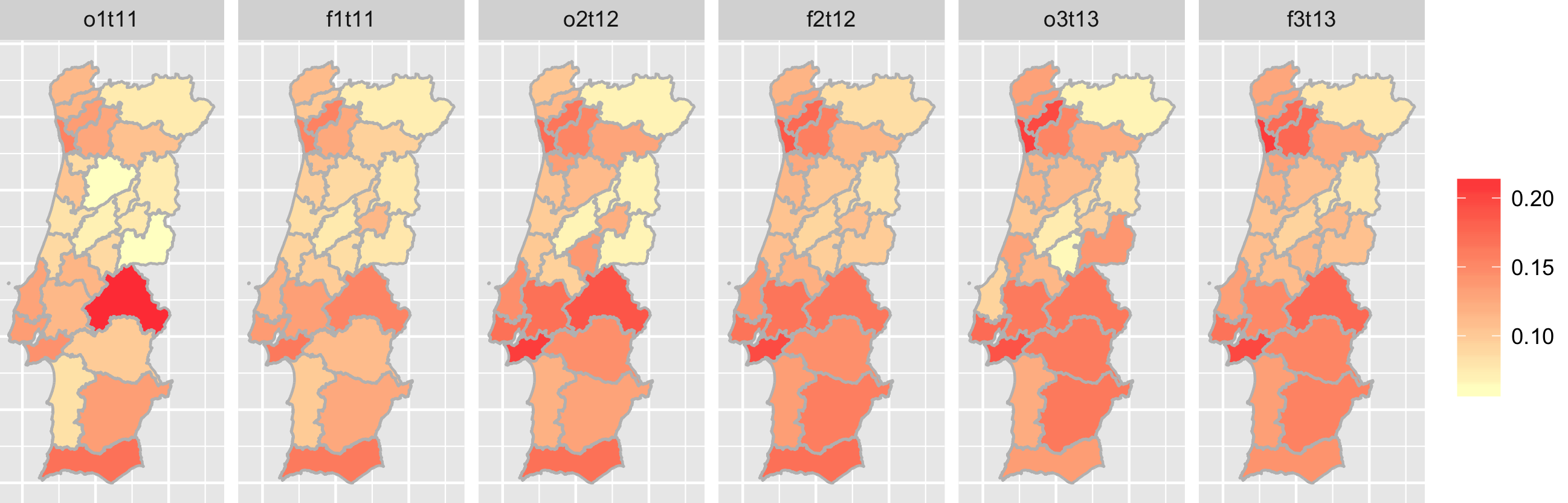}
\caption{Maps of observed and fitted values of the unemployment rate for the 1st quarter of 2011, 2nd quarter of 2012 and 3rd quarter of 2013.}
\label{mapas_binomial}
\end{figure}

The map of the figure \ref{mapas_binomial} allows for a better understanding of the regional difference between the observed and fitted values. 


\FloatBarrier

\subsection{Diagnosis}

Some predictive measures can be used for an informal diagnostic, such as \textit{Conditional Predictive ordinates} (CPO) and {Probability Integral Transforms} (PIT; Gelman et al, 2004). Measure $ CPO_i $ is defined as $ \pi(y_i | y_{-i}) $ where $ y_{-i} $ is the vector $ y $ without observation $ y_i $, while the measures $ PIT_i $ are obtained by $ Prob(y^{new}_i \le y_i | y_{- i}) $. Unusually large or small values of this measure indicate possible outliers. Moreover, a histogram of the PIT value which is very different from the uniform distribution indicates that the model is questionable.

The implementation of these measures in an MCMC approach is very heavy and requires a high computational time. For this reason, we present only results for the models implemented with the INLA.

\begin{figure}
\centering
\includegraphics[width=8cm]{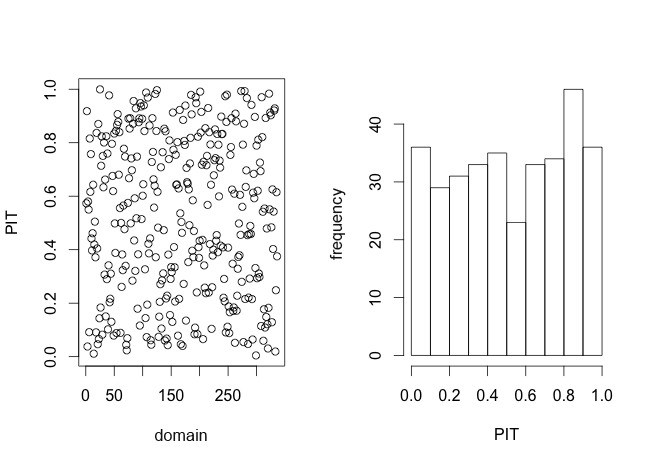}
\caption{Graphs of the PIT values versus domain ($ 28 \times 12 = $ 336) and the histogram of the PIT values.}
\label{pit}
\end{figure}

Figure \ref{pit} shows the graphs of the PIT values versus domain ($ 28 \times 12 = $ 336) and the histogram of the PIT values for Poisson, Binomial, Negative Binomial and Beta models. We see that the histogram for the PIT values based on the Poisson and Binomial models presents a fairly uniform behavior, but this is not the case with the Negative Binomial and Beta distributions. This suggests that these last two models may not be suitable for data in analysis.


The predictive quality of the models can be performed using a cross-validated logarithmic score given by the symmetric of the mean of the logarithm of CPO values (Martino and Rue, 2010). High CPO values indicate a better quality of prediction of the respective model. The logarithmic of the CPO values are given in table \ref{score}. Accordingly, the Beta model has the least predictive quality. 

The diagnosis of the Multinomial model was based on graphical visualization and on \textit{Potential Scale Reduction Factor} (Brooks and Gelman, 1997). No convergence problems were detected. 

\begin{table}
\captionsetup{width=0.8\linewidth}
\caption{Logarithmic score \label{score}}
\centering
{\begin{tabular}{lr}
\hline
&  log score \\
\hline
Poisson &	3.33 \\
Negative Binomial &	3.51 \\
Binomial &	3.34 \\
Beta &	-2.39	\\
\hline
\end{tabular}}
\end{table}

\FloatBarrier


\subsection{Comparison}

In order to compare the studied models, we use the \textit{Deviance Information Criterion} ($DIC$) proposed by Spiegehalter \textit{et al} (2002). This is a criterion which aims to achieve a balance between the adequacy of a model and its complexity. It is defined by $ DIC = \bar{D} + p_D $ where $ \bar{D} $ is the posterior mean deviance of the model and $ p_D $ is the effective number of parameters. The model with the smallest value of DIC is the one with a better balance between the model adjustment and complexity.

The values of $ DIC$, $ p_D $ and $\bar{D}$ are presented in table \ref{DIC}. The Multinomial model features a higher DIC value, but it should be noted that this model requires an adjustment of the total of employed, unemployed and inactive people, unlike the other models.

Among the models used for modeling of total, the Poisson model is the one with the lower value of DIC, which would suggest that it should  be preferable to the Negative Binomial model. However, Geedipally \textit{et al} (2013) explains that the value of this measure is affected by the parameterization of the model, which may influence the values obtained by the Negative Binomial and Beta models, since the software used permits different parameterizations in these cases.

\begin{table}
\captionsetup{width=0.8\linewidth}
\caption{DIC, effective number of parameters, and posterior mean of the deviance. \label{DIC}}
\centering
{\begin{tabular}{lrrr}
\hline
&  $DIC$ & \textbf{$p_D$} & \textbf{$\bar{D}$} \\
\hline
Poisson &	2240.4 &  30.4 & 2210.0 \\
Negative Binomial &	2374.9 &  25.4 &2349.5 \\
Binomial &	2241.4 &  32.5 &2208.9 \\
Beta &	-1607.4 &   31.1 & -1638.4	\\
Multinomial & 4976.0  & 81.5  & 4894.5 \\
\hline
\end{tabular}}
\end{table}

\begin{figure}[!htbp]
\begin{subfigure}{.5\textwidth}
  \centering
  \includegraphics[width=1\linewidth]{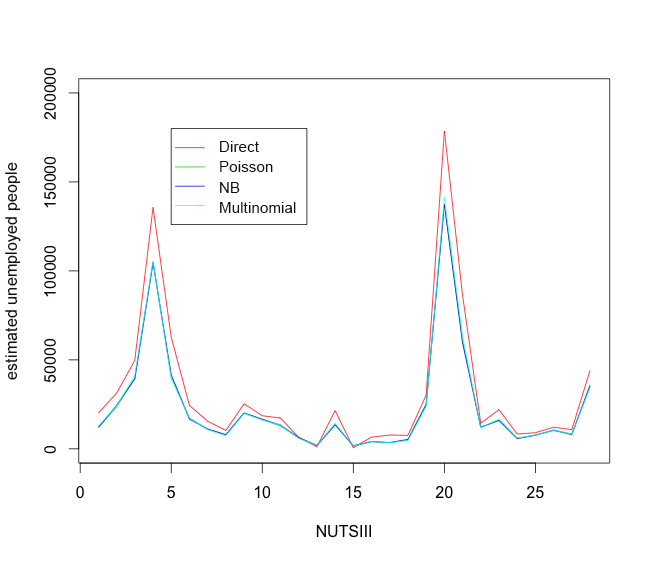}
  \caption{Totals}
  \label{fig:sfig1}
\end{subfigure}%
\begin{subfigure}{.5\textwidth}
  \centering
  \includegraphics[width=1\linewidth]{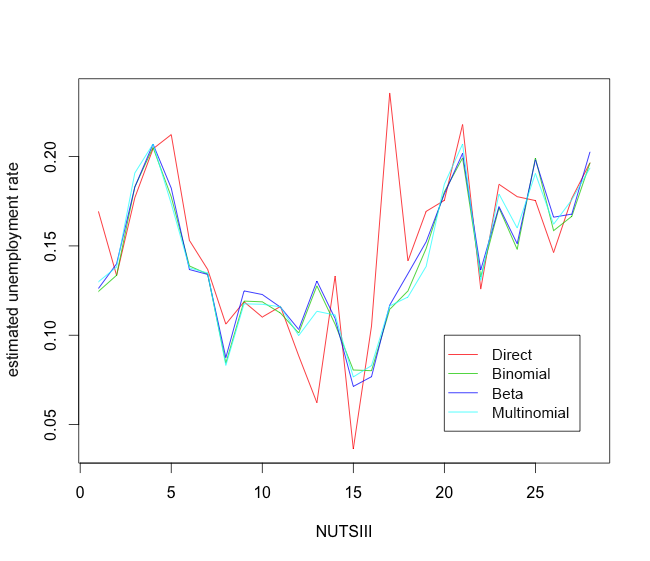}
  \caption{Rates}
  \label{fig:sfig2}
\end{subfigure}
\caption{Estimates for the total unemployed (through Poisson, Negative Binomial, Multinomial, and direct method) and the unemployment rate (through Binomial, Beta, Multinomial, and direct method).}
\label{totais_taxas}
\end{figure}

Figure  \ref{totais_taxas} shows that the Poisson, Negative Binomial and Multinomial models produced very similar estimates for the total unemployed, while Binomial, Beta and Multinomial models produced similar estimates for unemployment rate. We can also note that these estimates are smoother than the estimates obtained by the direct method. This property is prominently displayed in the estimation of the unemployment rate, and justifies the fact that the estimates of the total produced by the models are lower than the estimates produced by the direct method for large values of unemployment, and higher for small values (regions 13 and 15).

Regions 4 and 20, which correspond to Grande Porto and Grande Lisboa, are those with the highest population size. This fact explains the high values of unemployment totals. On the other hand, regions 13 and 15, which correspond to Pinhal Interior Sul and Serra da Estrela, are those with the lowest population size. It is interesting to observe that the regions with the greatest difference between the unemployment rate estimated by the direct method and the rate estimated by the studied models are those with the lowest population sizes (Pinhal Interior Sul, Serra da Estrela and Beira Interior Sul), which are represented in the graph by the numbers 13, 15 and 17.


\begin{figure}[!htbp]
\begin{subfigure}{.5\textwidth}
  \centering
  \includegraphics[width=1\linewidth]{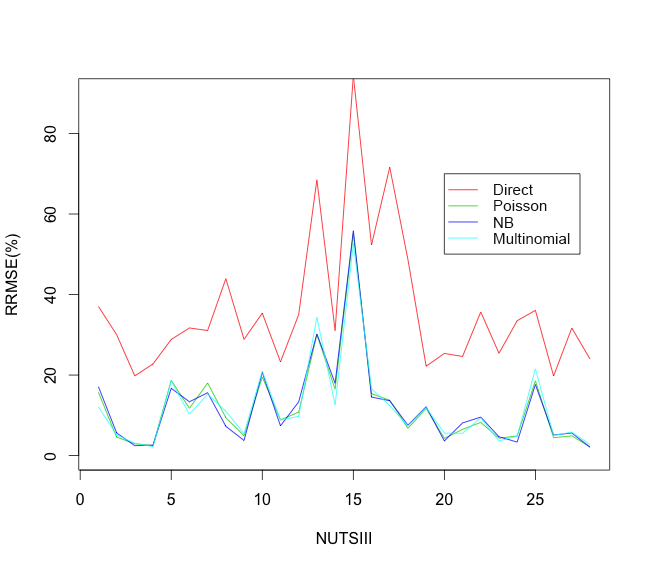}
  \caption{Totals}
  \label{fig:sfig1}
\end{subfigure}%
\begin{subfigure}{.5\textwidth}
  \centering
  \includegraphics[width=1\linewidth]{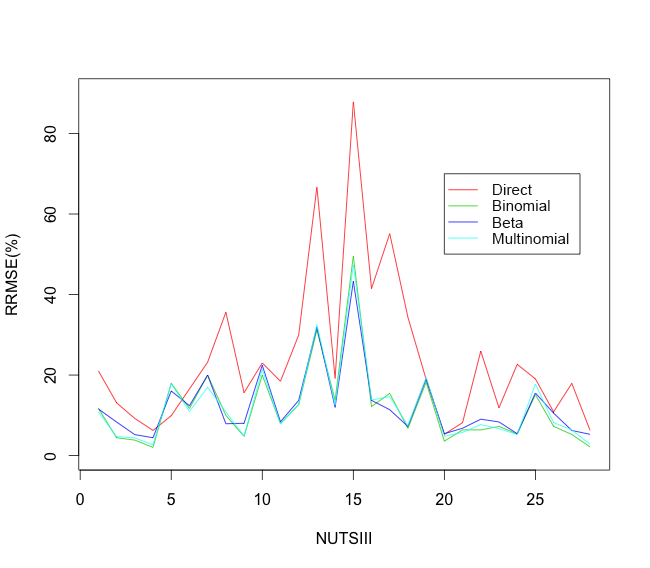}
  \caption{Rates}
  \label{fig:sfig2}
\end{subfigure}
\caption{RRMSE estimates for the total unemployed and the unemployment rate.}
\label{rrmse}
\end{figure}

The Relative Root Mean Square Error (RRMSE) allows for a comparison of the models studied and the direct method. A lower value of RRMSE indicates a better balance between variability and bias. The graph of Figure \ref{rrmse}  reveals a wide discrepancy between the direct method and the applied models.

Note that, for most models, the NUTS III region 15, which corresponds to Serra da Estrela (see Appendix), presents the highest value RRMSE. This result can be explained in part by the reduced population size of the region. The opposite is true for regions with high dimensional population such as Porto (Region 4), Grande Lisboa (region 20) and Algarve (region 28).

The high values of RRMSE of unemployment rate estimates by the direct method for regions 13, 15 and 17, can explain the big differences found between the methods in these regions (figure 10 b).
These results reinforce the idea that the direct method is inadequate for the estimation in small areas. On the other hand, the models studied show a maximum value of RRMSE of the unemployment rate estimates that corresponds to almost half of the value obtained by the direct method.


\FloatBarrier

\section{Discussion}

We studied the application of five spatio-temporal models within a bayesian approach for the estimation of both the total and the rate of unemployment of NUTS III regions.
We realized that one of the features of model based methods is the smoothing of the variation across time and space. This feature brings these models closer to reality.


The estimates obtained by these models were reasonable when compared with the direct method, which presented higher values of RRMSE.

Models under study presented much lower values of RRMSE than the direct method for regions with a small population size. This feature shows that these models can be a good alternative to small area estimation and in particular for the NUTS III regions of Portugal. 

The Negative Binomial and Beta models presented diagnostic problems in the analysis of empirical distribution of the PIT. A non uniform distribution of the PIT revealed that the predictive distribution is not coherent with the data.

Among the models under study, the Multinomial model seems to be the most suitable for this problem. The estimates obtained for the unemployment totals are similar to those obtained by the other models, but they produce estimates for the total of employed as well as inactive people simultaneously, in a way that is consistent with the population estimates. In this way, we can directly obtain the estimates of the employment and unemployment rates.

\FloatBarrier

\section*{Acknowledgements}

This work was supported by the project UID/MAT/00006/2013 and the PhD scholarship SFRH/BD/92728/2013 from Fundação para a Ciência e Tecnologia. Instituto Nacional de Estatística and Centro de Estatística e Aplicações da Universidade de Lisboa are the reception institutions. We would like to thank professor Antónia Turkman for her help.

\FloatBarrier

\section*{Note}

This study is the responsibility of the authors and does not reflect the official opinions of Instituto Nacional de Estatística.

\FloatBarrier


\section*{Appendix}
\begin{table}[!htbp]
\caption{NUTS II and NUTS III regions of Continental Portugal. \label{nuts}}
\centering
{\begin{tabular}{lllc}
\hline
& \multicolumn{2}{c}{NUTS III region}\\
\cline{2-3}
& Code &Designation &NUTS II region\\
\hline
1 &111 &Minho-Lima & Norte \\
2 &112 &Cávado &\\
3 &113 &Ave &\\
4 &114 &Grande Porto &\\
5 &115 &Tâmega &\\
6 &116 &Entre Douro e Vouga &\\
7 &117 &Douro &\\
8 &118 &Alto Trás-os-Montes &\\
\hline
9 &161 &Baixo Vouga & Centro \\
10 &162 &Baixo Mondego &\\
11 &163 &Pinhal Litoral  &\\
12 &164 &Pinhal Interior Norte &\\
13 &166 &Pinhal Interior Sul &\\
14 &165 &Dão-Lafões &\\
15 &167 &Serra da Estrela  &\\
16 &168 &Beira Interior Norte  &\\
17 &169 &Beira Interior Sul  &\\
18 &16A &Cova da Beira  &\\
19 &16B &Oeste &\\
\hline
20 &171 &Grande Lisboa  &  Lisboa\\
21 &172 &Península de Setúbal  &\\
\hline
22 &16C &Médio Tejo   &Centro\\
\hline
23 &185 &Lezíria do Tejo  & Alentejo\\
24 &181 &Alentejo Litoral  &\\
25 &182 &Alto Alentejo  &\\
26 &183 &Alentejo Central  &\\
27 &184 &Baixo Alentejo  &\\
\hline
28 &150 &Algarve &Algarve \\
\hline
\end{tabular}}
\end{table}


\FloatBarrier

\end{document}